\documentstyle[prb,aps,psfig]{revtex}
\begin{document}

\def\NPB{{\sf Nucl. Phys. }{\bf B}}
\def\PL{{\sf Phys. Lett. }}
\def\PRL{{\sf Phys. Rev. Lett. }}
\def\PRD{{\sf Phys. Rev. }{\bf D}}
\def\CQG{{\sf Class. Quantum Grav. }}
\def\JMP{{\sf J. Math. Phys. }}
\def\SJNP{{\sf Sov. J. Nucl. Phys. }}
\def\SPJ{{\sf Sov. Phys. J. }}
\def\JETPL{{\sf JETP Lett. }}
\def\TMP{{\sf Theor. Math. Phys. }}
\def\IJMPA{{\sf Int. J. Mod. Phys. }{\bf A}}
\def\MPL{{\sf Mod. Phys. Lett. }}
\def\CMP{{\sf Commun. Math. Phys. }}
\def\AP{{\sf Ann. Phys. }}
\def\PR{{\sf Phys. Rep. }}
\def\pl#1#2#3{Phys.~Lett.~{\bf {#1}B} (19{#2}) #3}
\def\np#1#2#3{Nucl.~Phys.~{\bf B{#1}} (19{#2}) #3}
\def\prl#1#2#3{Phys.~Rev.~Lett.~{\bf #1} (19{#2}) #3}
\def\pr#1#2#3{Phys.~Rev.~{\bf D{#1}} (19{#2}) #3}
\def\cqg#1#2#3{Class.~and Quantum Grav.~{\bf {#1}} (19{#2}) #3}
\def\cmp#1#2#3{Commun.~Math.~Phys.~{\bf {#1}} (19{#2}) #3}
\def\jmp#1#2#3{J.~Math.~Phys.~{\bf {#1}} (19{#2}) #3}
\def\ap#1#2#3{Ann.~of Phys.~{\bf {#1}} (19{#2}) #3}
\def\prep#1#2#3{Phys.~Rep.~{\bf {#1}C} (19{#2}) #3}
\def\ptp#1#2#3{Progr.~Theor.~Phys.~{\bf {#1}} (19{#2}) #3}
\def\ijmp#1#2#3{Int.~J.~Mod.~Phys.~{\bf A{#1}} (19{#2}) #3}
\def\mpl#1#2#3{Mod.~Phys.~Lett.~{\bf A{#1}} (19{#2}) #3}
\def\nc#1#2#3{Nuovo Cim.~{\bf {#1}} (19{#2}) #3}
\def\ibid#1#2#3{{\it ibid.}~{\bf {#1}} (19{#2}) #3}
\newcommand{\R}{{\bf R}}
\newcommand{\eq}{\begin{equation}}
\newcommand{\en}{\end{equation}}
\newcommand{\eqn}{\begin{eqnarray}}
\newcommand{\enn}{\end{eqnarray}}
\newcommand{\nn}{\nonumber }
\newcommand{\beq}{\begin{equation}}
\newcommand{\eeq}{\end{equation}}
\newcommand{\lb} {\langle}
\newcommand{\rb} {\rangle}

\draft
\twocolumn[\hsize\textwidth\columnwidth\hsize\csname @twocolumnfalse\endcsname
%
%
%

\title{Redistribution of Spectral Weight in Spin-1/2-Doped
Haldane Chains}             

\author{Stefan Wessel and Stephan Haas}
\address{Department of Physics and Astronomy, University of Southern
California, Los Angeles, CA 90089-0484}

\date{\today}
\maketitle

\begin{abstract}
We study the evolution of the dynamical spin structure factor in
a spin-1 antiferromagnetic Heisenberg chain which is randomly doped
with spin-1/2 moments. Using stochastic series expansion Quantum 
Monte Carlo simulations combined with the Maximum Entropy method,
we monitor the  crossover
from the spin-1 chain, dominated by 
a single resonance at the Haldane energy, to the spin-1/2 chain
with a continuum of spinon states which diverges at zero-frequency. 
Upon increasing the doping level, spectral weight is rapidly transferred
from the Haldane peak to lower energies. 
If the exchange couplings between the spin-1/2
substituents and the spin-1 sites of the host are sufficiently small,      
finite-frequency bound states are observed below the spin gap 
for small doping concentrations.  
\end{abstract}
\pacs{}
]

It is well established that antiferromagnetic (AF) spin-1 Heisenberg 
chains have a low-energy spin gap $\Delta \simeq 0.41J$, leading to
properties significantly different from quasi-long-range-ordered
spin-1/2 AF Heisenberg chains.
\cite{haldane,takahashi,golinelli,haas}
The effects of non-magnetic impurities on such
low-dimensional quantum spin liquids
have been the focus of considerable recent attention. 
In particular, it was shown
that open ends in segmented Haldane chains induce low-energy states 
below the spin gap,\cite{aklt} and that vacancies in AF
Heisenberg ladders liberate free spin degrees of freedom
in their immediate vicinity\cite{laukamp,martins}.
At sufficiently low energies, RKKY-like interactions among these
``pruned" spins lead to the formation of networks described
by effective models of random spins interacting via 
random exchange couplings.\cite{sigrist,wessel}
The effects of substitutions with non-vanishing magnetic moments 
tend to be more complex, depending on the spins of the dopants as
well as on their exchange couplings with the host material. 
\cite{sorensenaffleck,kaburagi} 
We address this issue
by studying the evolution of the dynamical spin structure factor
of spin-1 Heisenberg chains when doped with static spin-1/2 moments,
and by monitoring the crossover from a completely gapped
Haldane spectrum to that of a spin-1/2 chain upon complete 
substitution.

There have 
been several experimental and theoretical studies of the 
effects of spin-1/2 impurities in the spin-1 chain.
For example,
the substitution of $\rm{Cu^{2+}}$ for $\rm{Ni^{2+}}$ in the 
compound $\rm{Ni(C_2H_8N_2)NO_2(ClO_4)}$ (NENP) introduces spin-1/2 impurities 
on the original Ni sites, which have a weak ferromagnetic coupling to the $S=1$ 
chain-end spins.\cite{exp7a} On the other hand, substituting Ca$^{2+}$ for 
Y$^{3+}$ in 
$\rm{Y_2BaNiO_5}$  introduces carriers in the chains which become localized 
on the oxygen sites between
the $\rm{Ni^{2+}}$ ions. This breaks the superexchange path and replaces it 
by a direct exchange coupling to the spin-1/2 impurities.\cite{exp7b} In this
latter compound,
inelastic neutron scattering experiments found a substantial increase of the 
spectral function below the Haldane gap, indicating the creation of states 
below the energy of the spin gap.\cite{exp7b} 

Numerical diagonalization and density matrix renormalization group (DMRG)
techniques were used to study the emergence of
such impurity-induced low-energy states in 
the Heisenberg model.\cite{sorensenaffleck,kaburagi} The main
conclusion of these studies is that localized excited states exist below
the Haldane gap as long as
the exchange couplings $J'$ between the spin-1/2 and spin-1 sites 
are small compared to the couplings $J$
between the original spin-1 sites $J$, namely
if $J'/J \alt 0.45$. Other works have considered mobile holes.
\cite{dagottodynamics,batistaaligia} In addition to enhanced 
low-energy spectral weight in the doped compound, 
these studies also observe low-temperature spin-glass 
behavior, due to impurity-induced frustrating 
ferromagnetic inter-chain interactions.\cite{janodpayen} 

\begin{figure}
\centerline{\psfig{figure=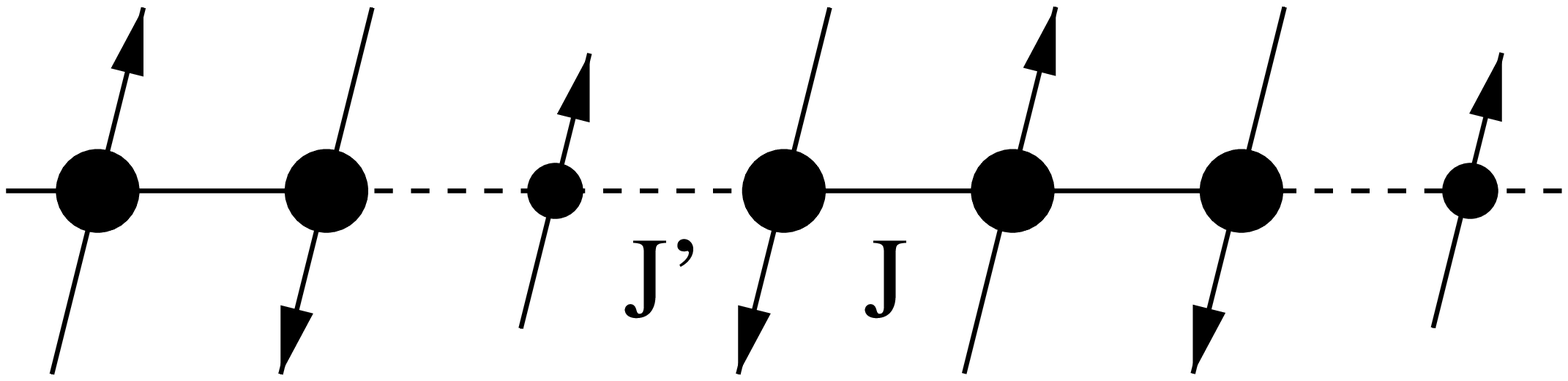,width=8cm,angle=0}}
\vspace{0.5cm}
\caption{Spin-1 AF Heisenberg chain with spin-1/2 moments. 
Exchange couplings between spin-1 sites are indicated by solid lines,
whereas the bonds between
spin-1/2 and spin-1 sites are denoted by dashed lines.}
\label{spec70}
\end{figure}

Here, we concentrate on the static case discussed in Refs. 
\cite{sorensenaffleck,kaburagi} and apply stochastic series expansion Monte 
Carlo  (SSE QMC) simulations\cite{sandvik}
in combination with the maximum entropy method\cite{deisz} to obtain
the spectral properties of Haldane chains randomly doped with spin-1/2 sites. 
This approach allows us to study the evolution
of the dynamical spin structure factor upon increasing the spin-1/2 
concentration far beyond the limit of low impurity concentration.
The doped spin-1 AF Heisenberg chain pictured in Fig. 1 is
modeled by a mixed-spin Heisenberg model,
\eq\label{mixham}
H=\sum_{i=1}^N J_i {\bf S}_i \cdot {\bf S}_{i+1},
\en
where ${\bf S}_i$ denotes a spin-1 or spin-1/2 moment. 
In general, three inequivalent types of bonds exist in a 
mixed-spin
chain: (i) bonds between two spin-1 sites, (ii) bonds between
two spin-1/2 sites, and (iii) bonds  
between spin-1 and spin-1/2 sites. 
In the mixed spin model of Eq. 1, we use
$J$ as the original coupling of the Haldane chain with
$J_i=J$ if $S_i$ and $S_{i+1}$ are both spin-1,
and otherwise $J_i=J'$, such that the
couplings between spin-1 and spin-1/2 sites 
are $J'$.
Since we are interested in the generic 
properties of
this problem, in our simulations we concentrate on representative
parameter sets.

The significance of $J'$ has been emphasized in Ref. \cite{sorensenaffleck} for
the case of isolated spin-1/2 impurities.
Consider first
the two extreme limits 
$J'/J = 0$ and $J'/J = \infty$. If the
spin-1/2 dopants are only weakly coupled to the spin-1 moments of the 
host, an effective 
Hamiltonian for the three-site cluster formed by the spin-1 chain-end spins and 
the impurity spin can be derived, correctly describing the emergence of 
bound states below the Haldane gap of the spin-1 chain.\cite{sorensenaffleck}
In the strong coupling limit,
the coupling to the remainder of the chain destabilizes the local
excitations. A critical coupling ratio $(J'/J)_{crit} = 0.45$ 
separates these two regimes. 
\cite{sorensenaffleck,kaburagi} 
Therefore, in the following we discuss separately the cases of
weak coupling, which we illustrate with the parameter choice
$J'=0.1 J$, and strong coupling, for which we take $J'=J$.

\begin{figure}
\centerline{\psfig{figure=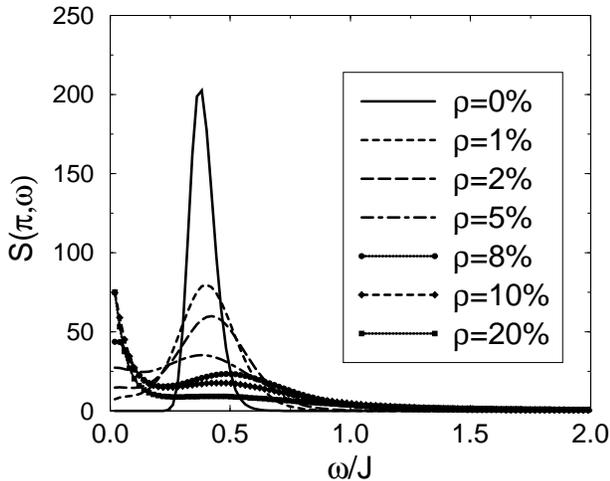,width=8cm,angle=0}}
\caption{Spectral function $S(\pi,\omega)$ of the spin-1 chain
doped with static 
spin-1/2 moments for various doping concentration $\rho$. The 
exchange couplings between spin-1 and spin-1/2 sites
are $J'=0.1J$ (weak-coupling limit).}
\label{spec71}
\end{figure}

In order to observe the evolution of the dynamical spectrum
with doping concentration, we have performed 
SSE QMC
simulations\cite{sandvik}
and measured the imaginary-time 
spin-spin correlation function
\eq\label{defS}
S({\mathbf{k}},\tau)=\frac{1}{N}\sum_{i,j} \exp{[-i\mathbf{k}(\mathbf{r}_i-
\mathbf{r}_j)]} \lb S_i^z(\tau)S^z_j(0) \rb,
\en
considering the dominant AF wave vector
$\mathbf{k}=\pi$ by using the Monte Carlo estimator 
derived in Ref. \cite{sandvikkurki}. 
We studied chains of up to 500 sites for up to 500 realizations of the  
random distribution of spin-1/2 substituents
down to temperatures 
$T\agt 0.01 J$. 
The maximum entropy analytical continuation procedure
was used
to obtain the spectral function $S(\pi,\omega)$ from the impurity-averaged data
for the imaginary-time correlation functions.\cite{deisz}

Let us first consider the case $J'/J=0.1$, where the emergence of localized 
states below the spin gap is expected from DMRG calculations in the single
impurity problem.\cite{sorensenaffleck,kaburagi} In Fig. \ref{spec71} we show 
the evolution of the spectral function  $S(\pi,\omega)$ upon increasing the 
concentration $\rho$ of spin-1/2 dopants. 
For the pure spin-1 chain, $\rho=0$, a dominant resonance is observed
at the Haldane gap energy 
$\omega=\Delta=0.41 J$, consistent with earlier results using 
similar methods.\cite{deisz} Upon introducing spin-1/2 sites,
the spectral
weight of the Haldane peak is strongly suppressed and its width broadens. In 
addition,
a significant amount of spectral weight appears at low but finite energies,
indicating the emergence of impurity-induced states below the gap edge. In 
fact, the shape of the spectral density for $\rho=0.01$ is consistent with 
Ref. \cite{sorensenaffleck}, i.e. bound states with energies of 
approximately $\alpha J'/2$ and $\alpha 3 J'/2$ appear in the low-energy 
spectrum, where $\alpha=1.064$ is a renormalization factor for the effective 
spin-1/2 chain-end spins ${\bf S}_{L,R}$ next to the impurity site.
\cite{white} These 
localized  states are the lowest excited states of the effective Hamiltonian 
$H_{eff}=\alpha J' ({\bf S}_L \cdot {\bf S}' + {\bf S}' \cdot {\bf S}_R)$, 
where ${\bf S}'$ denotes the impurity spin.\cite{sorensenaffleck} 
The maximum entropy method does not allow us to resolve the detailed structure 
of the spectrum, but clearly indicates the increase of spectral weight at 
this energy scale with increasing impurity concentration. Upon a further 
increase of the spin-1/2 concentration beyond $\rho \agt 0.05 $,
a zero-frequency peak emerges out of these low-energy contributions. 
Simultaneously, the Haldane peak appears to move towards larger frequencies
\cite{footnote}. 
The spectrum at larger impurity 
concentrations ($\rho>0.2$) exhibits features similar to those of 
the pure spin-1/2 chain, i.e., a divergence at $\omega=0$ and no remaining 
peak structure at the energy of the Haldane gap.

\begin{figure}
\centerline{\psfig{figure=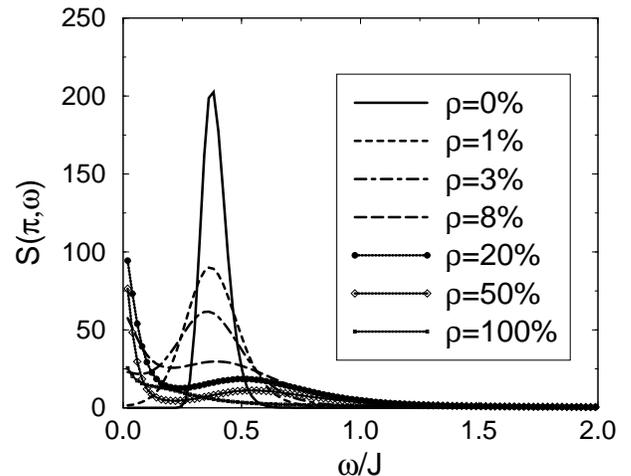,width=8cm,angle=0}}
\caption{Same as Fig. 2, but with isotropic exchange coupling constants,
$J'=J$, between all spin sites.}
\label{spec72}
\end{figure}

Turning to the isotropic case, $J'=J$, DMRG calculations for single
spin-1/2 impurities do not exhibit any 
low-energy states below the Haldane gap.\cite{sorensenaffleck} Instead, an 
extended state with an energy of the order of $\Delta$ is reported. We 
performed 
SSE QMC simulations for the isotropic case and 
obtained the dynamical structure factors shown in Fig. \ref{spec72}. 
Concentrating first on the low-energy properties of this
spectral function, we can confirm that
in the limit of small impurity concentrations $\rho<0.06$ 
no indications of low-lying impurity
states are observed, consistent with the DMRG results. 
Upon further increasing 
the impurity concentration, a peak appears at zero energy, which develops 
into the zero-frequency singularity characteristic of the gapless
spinon continuum 
of states in the pure spin-1/2 Heisenberg chain,
i.e. $S(\pi , \omega ) \propto \sqrt{\ln \omega}/\omega$.
Analogous to the discussion of the weak-coupling
limit, spectral weight is shifted
from the Haldane peak to lower energies. However,
in the isotropic case
no bound states are observed at finite frequencies
below the spin gap. Furthermore, the depletion of the Haldane resonance
is less rapid than in the weak-coupling limit, indicating that the 
presence of weaker bonds has a more dramatic effect on the redistribution 
of spectral weight than the presence of spin-1/2 substituents.  

\begin{figure}[t]
\centerline{\psfig{figure=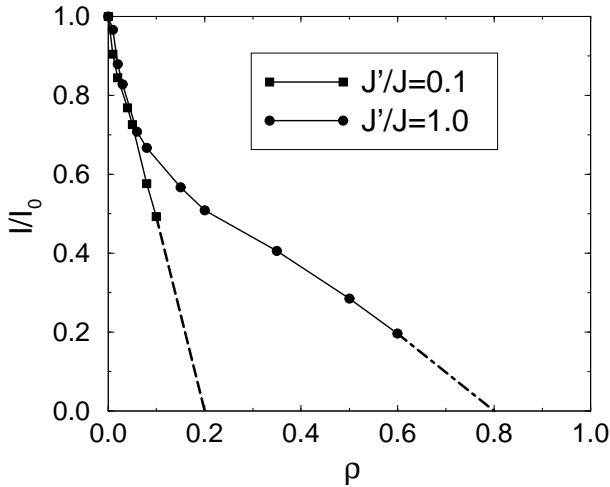,width=8cm,angle=0}}
\caption{Spectral weight $I$ of the 
Haldane peak, normalized to the spectral weight in the clean limit, $I_0$. 
The circles represent the isotropic case $J'=J$, whereas
the squares denote the weak-coupling limit
at $J'=0.1J$. The dashed  and dot-dashed
lines are linear extrapolations.} 
\label{fig73}
\end{figure}

In order to quantify this statement, 
the dependence of the spectral weight $I$ at the Haldane 
energy is plotted in Fig. \ref{fig73} as a function of the 
doping concentration. 
Here, a local Gaussian fit 
of the spectral function centered  around the peak position was used to 
estimate the spectral weight from the height and the width of the 
approximant. It is found that in
the isotropic case remnants of the Haldane peak can be 
resolved at much higher spin-1/2 concentrations
than in the limit of weak coupling. For example, at $\rho=0.5$, the 
Haldane resonance is
still visible in Fig. \ref{spec72} ($J' = J$), whereas in Fig.
\ref{spec71} ($J' = 0.1 J$) this feature 
cannot be resolved beyond $\rho=0.2$. This critical value for $J' = 0.1 J$ is 
obtained by a linear extrapolation of $I$ (dashed line shown in Fig. 
\ref{fig73}).
In the case of isotropic couplings the spectral weight of the 
Haldane peak initially shows a rapid decrease,
similar to the weak-coupling
limit. However, beyond a doping concentration
$\rho \agt 0.06$ at which zero-frequency singularities emerge in the 
dynamical structure factor (Fig. \ref{spec72}), $I$ decays 
at a significantly slower rate. 
It is obviously more difficult to deplete 
the extended states at the energy of the Haldane gap for
isotropic couplings than dissolving
the sub-gap bound states in the weak-coupling limit.
\cite{sorensenaffleck}
A linear extrapolation of $I$
(dot-dashed line in Fig. \ref{fig73}) suggests a critical doping concentration
$\rho\approx 0.8$ for isotropic couplings. 
Beyond this concentration the dynamical structure factor
resembles that of the pure spin-1/2 AF Heisenberg chain.
\cite{footnote2}

In conclusion, we have used a combination of stochastic series expansion 
Quantum Monte Carlo simulations and 
Maximum Entropy analytical continuation techniques
to study the effects of random spin-1/2 substitutions on the dynamical
response function of spin-1 AF Heisenberg
chains. In the limit of small doping concentrations,
the dynamical spin structure factor is found to agree very
well with results of earlier DMRG and exact diagonalization studies. 
If the AF couplings between the spin-1/2 dopants and the spin-1 sites
of the host are weak, bound states are formed
below the Haldane gap $\Delta \simeq 0.41 J$. On the other hand, 
in the case of isotropic coupling constants there are extended states at the 
energy of the spin gap. 
Upon increasing the doping concentration,
a proliferation of low-energy states is observed at the expense of the 
spectral weight of the dominant resonance at $\omega = \Delta$ of the pure
spin-1 AF Heisenberg chain.
In the limit of weak coupling, the 
spectral weight of the Haldane peak disappears rapidly. However, for isotropic
couplings the transfer of spectral weight to lower energies
is more gradual.
Here, the gapless spinon continuum spectrum of the pure spin-1/2 
AF Heisenberg chain is recovered
beyond a critical doping concentration $\rho\approx 0.8$.

We thank 
B. Normand
for useful discussions,
and acknowledge financial support by the Petroleum Research Fund, 
Grant No. ACS-PRF 35972-G6 and by the National Science Foundation, 
Grant No. DMR 0089882.


\begin{references}
\bibitem{footnote} Although the SSE QMC allows us to obtain imaginary-time
correlation functions with exceptionally small statistical errors, the 
Maximum Entropy method is known to introduce some inaccuracies in the 
position and widths of the peaks, in particular at higher frequencies.
\bibitem{footnote2} For $J' = J$ the signal of the Haldane peak is beyond
$\rho \approx 0.6$. While from the trend
of $I( \rho )$ a linear extrapolation yielding $\rho_c \approx 0.8$ 
appears to be reasonable, it is also possible that remnants of the 
Haldane peak exist up to $\rho \rightarrow 1$. Unfortunately, this
issue is beyond the resolution of the Maximum Entropy 
method used here.
\end{references}
\end{document}